\documentclass[12pt]{book}  
\usepackage{sussp}          
\usepackage{graphicx}     
\usepackage{makeidx}
\makeindex
\graphicspath{{./figs/}}

\def\plb#1#2#3{{\it Phys.\ Lett.} {\bf B#1}, #3 (20#2)}
\def\prd#1#2#3{{\it Phys.\ Rev.} {\bf D#1}, #3 (20#2)}

\def\prl#1#2#3{{\it Phys.\ Rev.\ Lett.} {\bf #1}, #3 (20#2)}

\def\sss{\scriptscriptstyle}	

\def\thingie{\hbox{\kern-9pt\raise1pt%
         \hbox{{\fiverm(}{\lower1.5pt\hbox{\twelvebf--}}{\fiverm)}}}}
\newcommand{\optbar}[1]{\shortstack{{\tiny (\rule[.4ex]{1em}{.1mm})} 
  \\ [-.7ex] $#1$}}		
\def\pmdiff#1#2{\raise.5ex\hbox{$\sss +#1}$%
    \kern-2.8em\lower1ex\hbox{${\sss-#2}$}} 
\def\pom{\raisebox{-.8ex}{$\stackrel{+}{{\sss (}-{\sss )}}$}}

\def\barp{{\raise.35ex\hbox{${\sss (}$}}---{\raise.35ex\hbox{${\sss )}$}}}		
\def\bdbarp{\hbox{$B_d$\kern-1.4em\raise1.4ex\hbox{\barp}}}
\def\nlpbarp{\hbox{$\nu_{\ell^{\prime}}$\kern-1.4em \raise1.4ex\hbox{\barp}}}
\def\decayarrow{\kern0.2em\hbox{$\raise1.08ex\hbox{\big|}\kern-0.5em \longrightarrow$}}

\def\gsim{\;\raisebox{-.6ex}{$\stackrel{>}{\sim}$}\;}
\def\ra{\rightarrow}

\def\dm2{\delta M^2_{ m \, m^{\sss\prime}}}
\newcommand{\Dmle}{\Delta m^2_{ij} L/2E}
\newcommand{\uai}{U_{\alpha i}}
\newcommand{\na}{\nu_\alpha}
\newcommand{\nb}{\nu_\beta}

\newcommand{\Eq}[1]{Eq.~(\ref{eq#1})}
\newcommand{\beeq}{\begin{equation}} 
\newcommand{\eeeq}{\end{equation}}

\begin{document}
\headings{Neutrino Oscillation Phenomenology$^{*\dag}$}
{Neutrino Oscillation Phenomenology}
{Boris Kayser}
{Fermilab, MS106, P.O. Box 500, Batavia IL 60510} 
\footnotetext{$^*$FERMILAB-PUB-08-079-T. To appear in the Proceedings of the 61st Scottish Universities Summer School in Physics: Neutrinos in Particle Physics, Astrophysics and Cosmology. Eds. C. Froggatt and P. Soler.}
\footnotetext{$^\dag$Author's e-mail address: boris@fnal.gov}

\section{Introduction}

Progress on our understanding of the neutrinos continues to be exhilarating. This progress is due mainly to experiments on neutrino oscillation. Here, we explain the physics of oscillation in vacuum and in matter.

\section{The physics of neutrino oscillation}

Treatments of the physics of neutrino oscillation may be found in, for example, [1, 2]. Here, we give a slightly modified treatment, and explain some points that have caused puzzlement, such as the fact that, even if neutrinos are their own antiparticles, their interaction with matter can still cause a difference between neutrino and so-called ``antineutrino'' oscillations.

We assume that the couplings of the neutrinos and charged leptons to the $W$ boson are correctly described by the Standard Model, extended to take leptonic mixing into account. These couplings are then summarized by the Lagrangian
\beeq
{\cal L}_W = -\frac{g}{\sqrt{2}} \sum_{{\stackrel{\alpha=e,\mu,\tau}{\sss i=1,2,3}}} (\overline{\ell_{L\alpha}} \gamma^\lambda U_{\alpha i} \nu_{Li} W_\lambda^- + \overline{\nu_{Li}} \gamma^\lambda U_{\alpha i}^* \ell_{L\alpha} W_\lambda^+) ~~.
\label{eq2.1}
\eeeq
Here, $L$ denotes left-handed chiral projection, $\ell_\alpha$ is the charged-lepton mass eigenstate of flavor $\alpha$ ($\ell_e$ is the electron, $\ell_\mu$ the muon, and $\ell_\tau$ the tau), and $\nu_i$ is a neutrino mass eigenstate. The constant $g$ is the semiweak coupling constant, and $U$ is the leptonic mixing matrix [3].
Supposing, as assumed by \Eq{2.1}, that there are only three charged-lepton mass eigenstates, and three neutrino mass eigenstates, $U$ is $3\times3$, and may be written as
\beeq
U = \left[ \begin{array}{ccc}
	U_{e1}      & U_{e2}       & U_{e3}        \\
	U_{\mu 1} & U_{\mu 2} & U_{\mu 3}    \\
	U_{\tau 1} & U_{\tau 2} & U_{\tau 3}
\end{array}\right]	~~ .
\label{eqI.3}
\eeeq

In the extended Standard Model, the $3\times 3$ mixing matrix $U$ is unitary, and we shall assume that this is also true in nature. However, we note that if there are ``sterile'' neutrinos (neutrinos that do not couple to the $W$ or $Z$ boson), then there are $N>3$ neutrino mass eigenstates, and the leptonic mixing matrix $U$ that is unitary is $N\times N$, rather than $3\times 3$. The $3\times 3$ matrix of \Eq{I.3} is then just a submatrix, and is not unitary [4].

Supposing that the unitary mixing matrix is $N\times N$, not because of the existence of sterile neutrinos but because there are $N$ conventional lepton generations, how many physically-significant parameters does $U$ contain? To see how many, we note first that an $N\times N$ complex matrix contains $N^2$ entries, each of which may have a real and an imaginary part. Thus, the matrix can be fully specified by $2N^2$ real parameters. If the matrix is unitary, then each of its columns must be a vector of unit length: $\sum_\alpha |\uai|^2 = 1; \;i=1,N$. Together, these conditions are $N$ constraints. In addition, each pair of columns in $U$ must be orthogonal vectors: $\sum_\alpha U_{\alpha i}^* \,U_{\alpha j} = 0; \; i,j=1,N$ with $i\neq j$. Taking into account that each of these $N(N-1)/2$ orthogonality conditions has both a real and an imaginary part, we see that these conditions impose $N(N-1)$ constraints. Thus, the number of independent parameters in a general $N\times N$ unitary matrix is $2N^2 -N -N(N-1) = N^2$. 
However, in the case of our unitary matrix, $U$, some of these parameters may be removed. From \Eq{2.1}, $\langle\ell_\alpha |{\cal L}_W| \nu_i W^-\!\rangle \;\propto \uai$. Now, without affecting the physics, we are always free to redefine the state $\langle\ell_\alpha|$ by multiplying it by a phase factor: $\langle\ell_\alpha| \ra \langle\ell_\alpha^\prime| = \langle\ell_\alpha| e^{-i \varphi_\alpha}$. Clearly, this has the effect of multiplying the $\uai$, for all $i$, by the same factor: $\uai \ra \uai^\prime = e^{-i \varphi_\alpha}\uai$. If there are $N \; \ell_\alpha$, this phase redefinition of them may be used to remove $N$ phases from $U$. It might be thought that analogous phase redefinition of the neutrinos $\nu_i$ could be used to remove additional phases. 
However, unlike the quarks and charged leptons, the neutrino mass eigenstates $\nu_i$ may be their own antiparticles: $\overline{\nu_i } = \nu_i$. This possibility motivates the search for neutrinoless nuclear double beta decay, as discussed at this school by K. Zuber. If $\overline{\nu_i } = \nu_i$, then physically significant phases cannot be eliminated by phase redefinition of the $\nu_i$ [5]. To allow for the possibility that $\overline{\nu_i } = \nu_i$, we shall retain the phases that can be eliminated only when $\overline{\nu_i } \neq \nu_i$. Then $U$ is left with $N^2 - N$ physically significant parameters. These are commonly chosen to be ``mixing angles''---parameters that would be present even if $U$ were real---and complex phase factors. To see how many of the parameters are mixing angles, and how many are phases, let us imagine for a moment that $U$ is real. 
Then it can be fully specified by its $N^2$ real entries. These are subject to the unitarity requirement that the N columns of $U$ all have unit length: $\sum_\alpha \uai^2 = 1,\; i=1,N$, and the requirement that all $N(N-1)/2$ pairs of columns be orthogonal: $\sum_\alpha \uai U_{\alpha j} = 0,\; i,j=1,N$ with $i\neq j$. Hence, a real mixing matrix $U$ for $N$ generations has $N^2-N-N(N-1)/2 = N(N-1)/2$ physically significant parameters, and a complex one has this number of mixing angles. Since a complex $U$ has $N(N-1)$ physically significant parameters in all, the fact that $N(N-1)/2$ of them are mixing angles means that the remaining $N(N-1)/2$ must be complex phase factors.

In summary, a complex $N\times N$ unitary mixing matrix $U$ for $N$ lepton generations may contain---
\begin{center}
\begin{tabular}{cc}
N(N-1)/2 & mixing angles  \\
N(N-1)/2& complex phase factors   \\
\hline
N(N-1) & physically significant parameters in all
\end{tabular}
\end{center}

Throughout most of these lecture notes, we will assume that $N=3$. Then the mixing matrix contains three mixing angles and three complex phase factors. It can be shown that this matrix can be written in the form
\begin{eqnarray}
U & = & \left[ \begin{array}{ccc}
	1    &      0      &    0       \\
	0    &  c_{23} & s_{23}  \\
	0    & -s_{23} & c_{23}
\end{array} \right]	\times	\left[ \begin{array}{ccc}
	 c_{13}	&	0	&	s_{13}e^{-i\delta}	\\
	 0		&	1	&	0				\\
	-s_{13}e^{i\delta}  & 0 &	c_{13}
\end{array} \right]	\times	\left[ \begin{array}{ccc}
	 c_{12}	& s_{12}	& 0	\\
	-s_{12}	& c_{12}	& 0	\\
	  0		&  0		& 1
\end{array} \right]	\nonumber \\
 &  \times	& \left[ \begin{array}{ccc}
	e^{i\xi_1 /2}	&	0	& 0	\\
	0		& e^{i\xi_2 /2}	& 0	\\
	0		&	0		& 1
\end{array} \right] ~~ .
\label{eqI.5.1}
\end{eqnarray}
Here, $c_{ij} \equiv \cos \theta_{ij}$ and $s_{ij} \equiv \sin \theta_{ij}$, where the $\theta_{ij}$ are the three mixing angles. The quantities $\delta,\; \xi_1$, and $\xi_2$ are the three complex phases.

From \Eq{2.1}, we observe that the amplitude for the decay $W^+ \ra \overline{\ell_\alpha} + \nu_i$ to yield the particular charged-lepton mass eigenstate $ \overline{\ell_\alpha}$ in combination with the particular neutrino mass eigenstate $\nu_i$ is proportional to $\uai^*$. Thus, if we define the ``neutrino state of flavor $\alpha$'', $|\nu_\alpha\rangle$, with $\alpha = e, \mu$, or $\tau$, to be the neutrino state that accompanies the particular charged lepton $ \overline{\ell_\alpha}$ in leptonic $W^+$ decay, then we must have
\beeq
|\nu_\alpha\rangle = \sum_{i=1}^3 \uai^* \, |\nu_i\rangle ~~ .
\label{eq2.2}
\eeeq
From \Eq{2.1}, the amplitude for this $\nu_\alpha$ to interact and produce the particular charged-lepton $\ell_\beta$ is proportional to 
\beeq
\sum_{i=1}^3 U_{\beta i} \uai^* = \delta_{\beta\alpha} ~~ ,
\label{eq2.3}
\eeeq
where we have invoked the unitarity of $U$. We see that when a $\nu_e$, the neutrino born in a $W^+$ decay that produced an $\bar{e}$, interacts and produces a second charged lepton, the latter can only be an $e$. Similarly for $\nu_\mu$ and $\nu_\tau$.

We may invert \Eq{2.2} to obtain
\beeq
|\nu_i\rangle = \sum_{\alpha = e, \mu, \tau} \uai |\nu_\alpha\rangle ~~ .
\label{eqI2.1}
\eeeq
This expresses the mass eigenstate $|\nu_i\rangle$ in terms of the states of definite flavor, $|\nu_\alpha\rangle$. We see that the flavor-$\alpha$ fraction of $|\nu_i\rangle$ is simply $|\uai|^2$.

\subsection{Neutrino oscillation in vacuum}

\begin{figure}[!htbp]
\centering
\includegraphics[width=14cm]{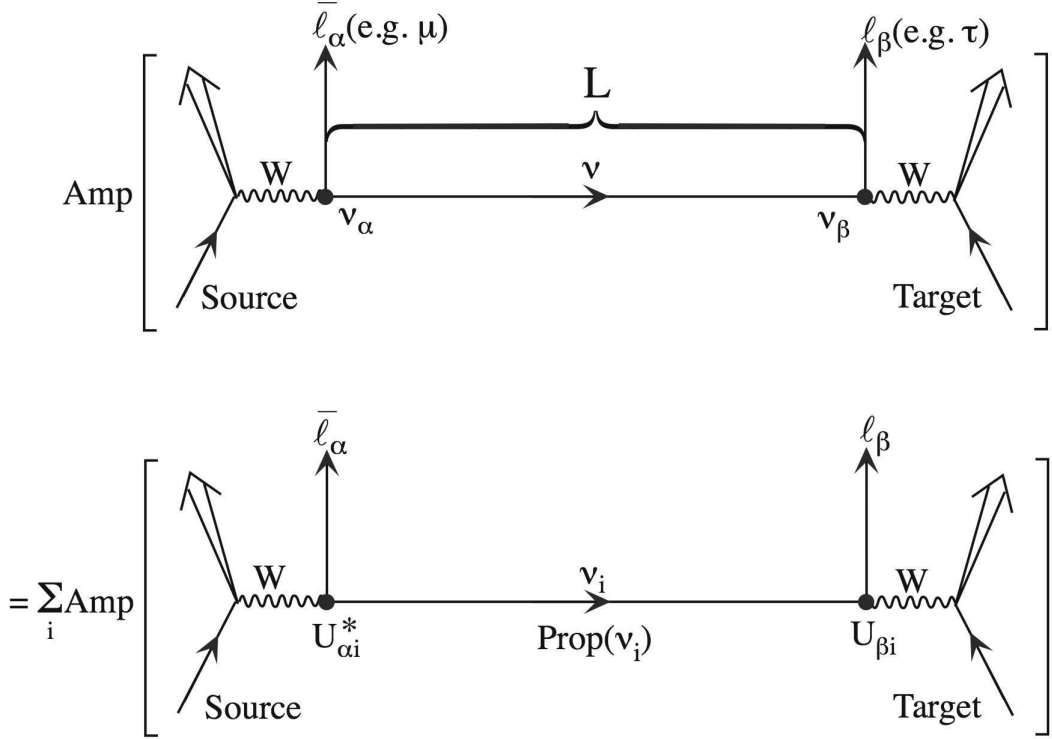}        
\caption{Neutrino flavor change (oscillation) in vacuum. ``Amp'' denotes an amplitude.} 
\label{f1}
\end{figure}

Consider the vacuum neutrino oscillation experiment depicted schematically in the upper part of Figure \ref{f1}.
A neutrino source produces, via $W$ exchange, the charged lepton $ \overline{\ell_\alpha}$ of flavor $\alpha$, plus an accompanying neutrino that, by definition, must be a $\na$. The neutrino then propagates, in vacuum, a distance $L$ to a target/detector. There, it interacts via $W$ exchange and produces a second charged lepton $\ell_\beta$ of flavor $\beta$. Thus, at the moment of its interaction in the detector, the neutrino is a $\nb$. If the flavors $\alpha$ and $\beta$ are different, then, during the neutrino's trip to the detector, it has changed, or ``oscillated'', from a $\na$ into a $\nb$.

In the neutrino mass eigenstate basis, the particle that travels from the neutrino source to the detector is one or another of the mass eigenstates $\nu_i$. In a given event, we will not know which $\nu_i$ was actually involved. Hence, the amplitude for the oscillation $\na \ra \nb$, Amp\,($\na \ra \nb$), is a coherent sum over the contributions of all the $\nu_i$, as shown in the lower part of Figure~\ref{f1}. The contribution of an individual $\nu_i$ is a product of three factors. The first is the amplitude for the neutrino produced together with the charged lepton $\overline{\ell_\alpha} $ to be, in particular, a $\nu_i$. From \Eq{2.1}, this amplitude is $\uai^*$, as indicated in Figure~\ref{f1}. The second factor, Prop\,($\nu_i$), is the amplitude for the mass eigenstate $\nu_i$ to propagate from the source to the detector. The final factor is the amplitude for the charged lepton created when the $\nu_i$ interacts in the detector to be, in particular, an $\ell_\beta$. From \Eq{2.1}, this amplitude is $U_{\beta i}$.

From elementary quantum mechanics, the propagation amplitude Prop($\nu_i$) is simply $\exp{[-im_i\tau_i}]$, where $m_i$ is the mass of $\nu_i$, and $\tau_i$ is the proper time that elapses in the $\nu_i$ rest frame during its propagation. By Lorentz invariance, $m_i\tau_i = E_i t - p_i L$, where $L$ is the lab-frame distance between the neutrino source and the detector, $t$ is the lab-frame time taken for the beam to traverse this distance, and $E_i$ and $p_i$ are, respectively, the lab-frame energy and momentum of the $\nu_i$ component of the beam.

Once the absolute square $|$Amp$\,(\na \ra \nb)|^2$ is taken to compute the probability for the oscillation $\na \ra \nb$, only the {\em relative} phases of the propagation amplitudes Prop\,($\nu_i$) for different mass eigenstates will have physical consequences. From the discussion above, the relative phase of Prop\,($\nu_1$) and Prop\,($\nu_2$), $\delta\phi(12)$ is given by
\begin{eqnarray}
\delta\phi(12)	& = & (E_2 t - p_2 L) - (E_1t-p_1L)  \nonumber  \\
			& = & (p_1 - p_2)L - (E_1 - E_2)t ~~ .
\label{eq2.4}
\end{eqnarray}
In practice, experiments do not measure the transit time $t$. However, Lipkin has shown [6] that, to an excellent approximation, $t$ may be taken to be $L/\bar{v}$, where
\beeq
\bar{v} \equiv \frac{p_1 + p_2}{E_1 + E_2}
\label{eq2.5}
\eeeq
is an approximation to the average of the velocities of the $\nu_1$ and $\nu_2$ components of the beam. We then have
\begin{eqnarray}
\delta\phi(12)	& \cong	& \frac{p_1^2 - p_2^2}{p_1 + p_2}L - \frac{E_1^2-E_2^2}{p_1 + p_2}L \nonumber   \\
			&   =		& (m_2^2-m_1^2)\frac{L}{p_1 + p_2} \cong (m_2^2-m_1^2)\frac{L}{2E} ~~ ,
\label{eq2.6}
\end{eqnarray}
where, in the last step, we have used the fact that for highly relativistic neutrinos, $p_1$ and $p_2$ are both approximately equal to the beam energy $E$. We conclude that all the relative phases in Amp($\,\na \ra \nb$) will be correct if we take
\beeq
\mathrm{Prop}\,(\nu_i) = e^{-im_i^2\,L/2E} ~~.
\label{eq2.7}
\eeeq

Combining the factors that appear in the lower part of Figure~\ref{f1}, we have 
\beeq
\mathrm{Amp}\,(\na \ra \nb) = \sum_i \uai^* e^{-im_i^2\,L/2E} U_{\beta i} ~~ .
\label{eq2.8}
\eeeq
Squaring, and making judicious use of the unitarity of $U$, we find that the probability of $\na \ra \nb$, P\,($\na\ra\nb$), is given by
\begin{eqnarray}
\mathrm{P}(\na\ra\nb)	& = &|\mathrm{Amp}\,(\na\ra\nb)|^2 \nonumber \\
		& = & \delta_{\alpha\beta} - 4\sum_{i>j} \Re (\uai^*U_{\beta i}U_{\alpha j}U_{\beta j}^*) \sin^2(\Delta m^2_{ij}L/4E)	\nonumber  \\
		&    & \phantom{ \delta_{\alpha\beta}}+ 2\sum_{i>j} \Im (\uai^*U_{\beta i}U_{\alpha j}U_{\beta j}^*) \sin(\Dmle) ~~ .
\label{eq2.9}
\end{eqnarray}
Here, $\Delta m^2_{ij} \equiv m^2_{i} - m^2_{j}$ is the splitting between the squared masses of $\nu_i$ and $\nu_j$. It is clear from the derivation of \Eq{2.9} that this expression would hold for any number of flavors and equal number of mass eigenstates.

Given that the particles described by the oscillation probability of  \Eq{2.9} are born with an $\overline{\ell_\alpha}$ and convert into an $\ell_\beta$ in the detector, they are {\em neutrinos}, rather than {\em antineutrinos} (should there be a difference). To obtain the corresponding oscillation probability for antineutrinos, we observe that $\overline{\na}\ra\overline{\nb}$ is the CPT-mirror image of $\nb\ra\na$. Thus, if CPT invariance holds,
\beeq
\mathrm{P}(\overline{\na}\ra\overline{\nb}) =  \mathrm{P}(\nb\ra\na) ~~ .
\label{eq2.10}
\eeeq
Now, from \Eq{2.9}, we see that
\beeq 
\mathrm{P}(\nb\ra\na;\; U) =  \mathrm{P}(\na\ra\nb;\; U^*) ~~ .
\label{eq2.11}
\eeeq
Hence, assuming CPT invariance holds,
\beeq
\mathrm{P}(\overline{\na}\ra\overline{\nb};\; U) = \mathrm{P}(\na\ra\nb;\; U^*) ~~ .
\label{eq2.12}
\eeeq
That is, the probability for oscillation of an antineutrino is the same as that for a neutrino, except that the mixing matrix $U$ is replaced by its complex conjugate. Thus, from \Eq{2.9},
\begin{eqnarray}
\mathrm{P}(\optbar{\na}\ra\optbar{\nb})	& = & \delta_{\alpha\beta} - 4\sum_{i>j} \Re (\uai^*U_{\beta i}U_{\alpha j}U_{\beta j}^*) \sin^2(\Delta m^2_{ij} L/4E)	\nonumber  \\
		&    &  \phantom{ \delta_{\alpha\beta}}\pom 2\sum_{i>j} \Im (\uai^* U_{\beta i}U_{\alpha j}U_{\beta j}^*) \sin(\Dmle) ~~ .
\label{eq2.13}
\end{eqnarray}
We see that if $U$ is not real, the probabilities for $\na\ra\nb$ and for the corresponding antineutrino oscillation, $\overline{\na}\ra\overline{\nb}$, will in general differ. Since  $\na\ra\nb$ and $\overline{\na} \ra \overline{\nb}$ are CP-mirror-image processes, this difference will be a violation of CP invariance.

As \Eq{2.13} makes clear, neutrino oscillation in vacuum from one flavor $\alpha$ into a different one $\beta$ implies nonzero mass splittings $\Delta m^2_{ij}$, hence nonzero neutrino masses. It also implies nontrivial leptonic mixing. That is, the mixing matrix $U$ cannot be diagonal. 

Including the so-far omitted factors of $\hbar$ and $c$, we have
\beeq
\Delta m^2_{ij} \frac{L}{4E} = 1.27 \, \Delta m^2_{ij} \mathrm{(eV}^2) \frac{L\mathrm{(km)}} {E\mathrm{(GeV)}} ~~ .
\label{eq2.14}
\eeeq
From \Eq{2.13}, if the $U$ matrix cooperates, the probability for $\na\ra\nb,\; \beta \neq \alpha$, will be appreciable if the kinematical phase difference in \Eq{2.14} is ${\cal O}(1)$ or larger. This requires only that for some $ij$,
\beeq
\Delta m^2_{ij} \mathrm{(eV}^2)\gsim \frac {E\mathrm{(GeV)}}{L\mathrm{(km)}} ~~ .
\label{eq2.15}
\eeeq
Thus, for example, an experiment that studies 1\,GeV neutrinos that travel a distance $L \sim 10^4$km, the diameter of the earth, will be sensitive to neutrino (mass)$^2$ splittings $\Delta m^2_{ij}$ as small as 10$^{-4}$eV$^2$. Through quantum interference between neutrino mass eigenstates of different masses, neutrino oscillation gives us sensitivity to very tiny (mass)$^2$ splittings. However, as \Eq{2.13} underscores, oscillation cannot determine the masses $m_i$ of the individual mass eigenstates. To learn those will require another approach. 

There are basically two kinds of neutrino oscillation experiments. In the first, an {\em appearance} experiment, one starts with a beam of neutrinos that initially are purely of flavor $\alpha$, and looks for the appearance in this beam of neutrinos of a new flavor $\beta,\; \beta \neq \alpha$, that were not originally present in the beam. In the second kind of experiment, a {\em disappearance} experiment, one starts with a known flux of $\na$, and looks to see whether some of the initial $\na$ flux disappears as the beam travels.

By the definition of ``probability'', the probability that a neutrino changes flavor, plus the probability that it does not change flavor, must equal unity. That is, we must have
\beeq
\sum_\beta \mathrm{P}(\na\ra\nb) =  \sum_\beta \mathrm{P}(\overline{\na}\ra\overline{\nb}) =1 ~~ ,
\label{eq2.16}
\eeeq
where the sum is over all final flavors $\beta$, including the initial flavor $\alpha$. From the unitarity of $U$, which implies that $\sum_\beta U_{\beta i}U_{\beta j}^* = \delta_{ij}$, it immediately follows that the oscillation probabilities of \Eq{2.13} do obey this constraint.

Neutrino flavor oscillation does not change the total flux in a neutrino beam. It merely redistributes it among the flavors. However, if we create a beam of neutrinos that at birth are of some active (i.e., weakly interacting) flavor, (muon neutrinos, for example), and some of these neutrinos oscillate into sterile (i.e., non-interacting) flavors, then some of the total {\em active} neutrino flux will have disappeared.

The combination of the CPT-invariance constraint of \Eq{2.10} and the probability constraint of \Eq{2.16} has powerful consequences for CP violation. To see this, consider the CP-violating differences
\beeq
\Delta_{\alpha\beta} \equiv \mathrm{P}(\na\ra\nb) -   \mathrm{P}(\overline{\na}\ra\overline{\nb}) ~~ .
\label{eq2.17}
\eeeq
If CPT invariance holds, then from \Eq{2.10}
\beeq
\Delta_{\beta\alpha} = -\Delta_{\alpha\beta} ~~ .
\label{eq2.18}
\eeeq
In particular,
\beeq
\Delta_{\alpha\alpha} = 0 ~~ .
\label{eq2.19}
\eeeq
That is, there can be no CP-violating difference between the survival probabilities P$(\na\ra\na)$ and P$(\overline{\na}\ra\overline{\na})$. Hence, there can be no observable CP violation in a disappearance experiment. Now, from \Eq{2.16}, it follows that
\beeq
\sum_\beta \Delta_{\alpha\beta} = 0 ~~ ,
\label{eq2.20}
\eeeq
where the sum runs over all flavors, including $\beta=\alpha$. However, in view of \Eq{2.19}, \Eq{2.20} implies that
\beeq
\sum_{\beta\neq\alpha} \Delta_{\alpha\beta} = 0 ~~ .
\label{eq2.21}
\eeeq
If there are only three neutrino flavors, $\nu_e, \;\nu_\mu$, and $\nu_\tau$, then this constraint implies that, in particular, 
\beeq
\Delta_{e\mu} + \Delta_{e\tau} = 0	\hspace{1.5cm}		\mathrm{and}	\hspace{1.5cm}
\Delta_{\mu e} + \Delta_{\mu\tau} = 0 ~~ .
\label{eq2.22}
\eeeq
From these relations and \Eq{2.18}, we see that
\beeq
\Delta_{e\mu} = \Delta_{\mu\tau} = \Delta_{\tau e} = -\Delta_{\mu e} = -\Delta_{\tau\mu} = -\Delta_{e\tau} \equiv \Delta ~~ .
\label{eq2.23}
\eeeq
In summary, if CPT holds, then the CP-violating difference $\Delta_{\alpha\beta} = \mathrm{P}(\na\ra\nb) - \mathrm{P}(\overline{\na}\ra\overline{\nb})$ can be nonvanishing only for $\beta \neq \alpha$. If, in addition, there are only three flavors, then the six possibly-nonvanishing $\Delta_{\alpha\beta}$, shown in \Eq{2.23}, must all be equal, apart from a predicted minus sign [7]. (If there are more than three flavors, then \Eq{2.23} need not hold.)

\begin{figure}[!htbp]
\centering
\includegraphics[width=14cm]{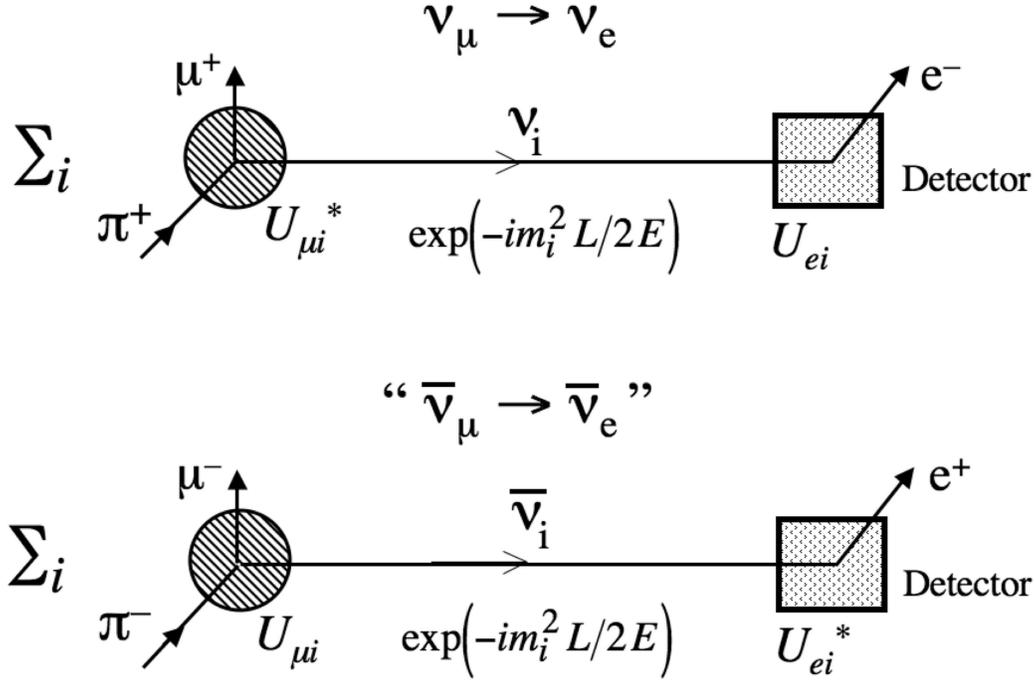}        
\caption{The CP-mirror-image oscillations $\nu_\mu \ra \nu_e$ and ``$\overline{\nu_\mu} \ra \overline{\nu_e}$''. In each process, the particle that travels down the beamline is one or another of the mass eigenstates, and the amplitude is a coherent sum over the contributions of these eigenstates, as indicated. In ``$\overline{\nu_\mu} \ra \overline{\nu_e}$'', the mass eigenstate $\overline{\nu_i}$ may or may not be identical, apart from its polarization, to the corresponding $\nu_i$ in $\nu_\mu \ra \nu_e$. The propagator for this $\overline{\nu_i}$, $\exp(-im^2_i L/2E)$, is identical to that for the corresponding $\nu_i$ in either case. The elements of the $U$ matrix that, according to \Eq{2.1}, appear at the beam-particle production and detection vertices are shown.}
\label{f1-1}
\end{figure}

Counter to intuition, the CP-violating difference $\Delta_{\alpha\beta} \equiv \mathrm{P}(\na\ra\nb) - \mathrm{P}(\overline{\na}\ra\overline{\nb})$ between neutrino and what we conventionally call ``antineutrino'' oscillation probabilities can still be nonvanishing even when the $\nu_i$ are identical to their antiparticles. 
Indeed, $\Delta_{\alpha\beta}$ is actually completely independent of whether the $\nu_i$ are their own antiparticles or not. We illustrate this by comparing the processes $\nu_\mu \ra \nu_e$ and
 ``$\overline{\nu_\mu} \ra \overline{\nu_e}$'', depicted in Figure~\ref{f1-1}.
In $\nu_\mu \ra \nu_e$, the neutrino is created together with a $\mu^+$ in $\pi^+$ decay. After traveling down a beamline to a detector, it is detected via its production of an $e^-$. In the corresponding ``antineutrino'' oscillation, ``$\overline{\nu_\mu} \ra \overline{\nu_e}$'', the particle that travels down the beamline is created together with a $\mu^-$ in $\pi^-$ decay, and is detected via its production in the detector of an $e^+$. 
One never directly observes the particle that travels down the beamline; it is an intermediate state. In terms of the charged leptons that one does ( or at least can) observe, $\nu_\mu \ra \nu_e$ and ``$\overline{\nu_\mu} \ra \overline{\nu_e}$'' are clearly different, CP-mirror-image processes: the first involves a $\mu^+$ and $e^-$, while the second involves a $\mu^-$ and $e^+$. 
Thus, even if $\overline{\nu_i} = \nu_i$, $\nu_\mu \ra \nu_e$ and ``$\overline{\nu_\mu} \ra \overline{\nu_e}$'' can have different probabilities, and if they do, the difference is a violation of CP invariance.

Even if $\overline{\nu_i} = \nu_i$, the beam particle will create an $e^-$ in $\nu_\mu \ra \nu_e$, but an $e^+$ in ``$\overline{\nu_\mu} \ra \overline{\nu_e}$'', because it is oppositely polarized in the two processes. Due to the chirally left-handed structure of the weak interaction, reflected in the Lagrangian of \Eq{2.1}, the beam particle will have helicity $h = -1/2$ in the first process, but $h = +1/2$ in the second. Due to this same parity-violating left-handed structure, the $h = -1/2$ beam particle will create an $e^-$ (via the first term in \Eq{2.1}) in $\nu_\mu \ra \nu_e$, while the $h = +1/2$ beam particle will create an $e^+$ (via the second term in \Eq{2.1}) in ``$\overline{\nu_\mu} \ra \overline{\nu_e}$''.

From the amplitude factors displayed in Figure~\ref{f1-1}, we see that while
\beeq
\mathrm{Amp} (\nu_\mu \ra \nu_e) = \sum_i U_{\mu i}^* e^{-im^2_i \frac{L}{2E}} U_{ei} ~~ ,
\eeeq
\beeq
\mathrm{Amp} (\overline{\nu_\mu} \ra \overline{\nu_e}) = \sum_i U_{\mu i} e^{-im^2_i \frac{L}{2E}} U_{ei}^*~~ .
\label{eq2.23b}
\eeeq
These expressions hold whether $\overline{\nu_i} = \nu_i$ or not. Thus, in either case, if the CP-violating phase $\delta$ in \Eq{I.5.1} is not zero or $\pi$, so that $U$ is complex, the interference terms in P($\nu_\mu \ra \nu_e$) and P($\overline{\nu_\mu} \ra \overline{\nu_e}$) will differ. As a result, the CP-violating difference P($\nu_\mu \ra \nu_e$) -- P($\overline{\nu_\mu} \ra \overline{\nu_e}$) will be nonzero. Furthermore, the value of this difference will not depend on whether $\overline{\nu_i} = \nu_i$, and this value will be correctly implied by \Eq{2.13}, which holds regardless of whether $\overline{\nu_i} = \nu_i$.

The general expression for P$(\optbar{\na} \ra \optbar{\nb})$, \Eq{2.13}, simplifies considerably in some important special cases. One such case is the simplified world in which there are only two charged leptons, say $e$ and $\mu$, two corresponding neutrinos of definite flavor, $\nu_e$ and $\nu_\mu$, and two neutrino mass eigenstates, $\nu_1$ and $\nu_2$, that make up $\nu_e$ and $\nu_\mu$. From our earlier analysis of the number of parameters in a mixing matrix, we know that the $2 \times 2$ unitary mixing matrix $U$ for this two-flavor world may contain one mixing angle and one complex phase factor. It may easily be shown that $U$ may be written in the form
\beeq
U \equiv \left[ \begin{array}{cc}
	U_{e1} & U_{e2}    \\    U_{\mu 1} & U_{\mu 2}
	\end{array} \right]    =
	\left[ \begin{array}{cc}
	\phantom{-}\cos\theta & \sin\theta  \\  -\sin\theta & \cos\theta
	\end{array} \right]    \times
	\left[ \begin{array}{cc}
	e^{i\xi /2} & 0    \\    0 & 1
	\end{array} \right] ~~,
\label{eqI.7}
\eeeq
where $\theta$ is the mixing angle and $\xi$ is the phase.
With $\Delta m^2_{21} \equiv \Delta m^2$ the sole (mass)$^2$ splitting in the problem, we find from the $U$ of \Eq{I.7} and the general expression of \Eq{2.13} that
\beeq
\mathrm{P}(\optbar{\nu_e}\ra\optbar{\nu_\mu})	 = \mathrm{P}(\optbar{\nu_\mu}\ra\optbar{\nu_e}) =   \sin^2 2\theta \sin^2(\Delta m^2 L/4E) ~~ ,
\label{eq2.24}
\eeeq
and that
\beeq
\mathrm{P}(\optbar{\nu_e}\ra\optbar{\nu_e}) = \mathrm{P}(\optbar{\nu_\mu}\ra\optbar{\nu_\mu}) =   1 - \sin^2 2\theta \sin^2(\Delta m^2 L/4E) ~~ .
\label{eq2.25}
\eeeq

\begin{figure}[!htbp]
\centering
\includegraphics[width=14cm]{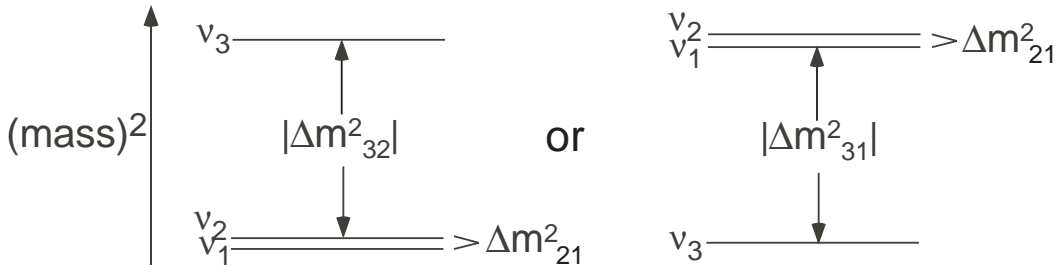}        
\caption{The three-neutrino (mass)$^2$ spectrum.} 
\label{f2}
\end{figure}

As we know, the real world contains (at least) three charged leptons $\ell_\alpha$, three corresponding neutrinos of definite flavor $\na$, and three underlying neutrino mass eigenstates $\nu_i$ that make up the $\na$. Thus, the two-neutrino oscillation formulae of Eqs.\ (\ref{eq2.24}) and  (\ref{eq2.25}) do not apply. However, if there are only three flavors, then under certain circumstances, rather similar simple formulae do apply. To see this, we note that the three-neutrino (mass)$^2$ spectrum has been observed to have the form shown in Figure~\ref{f2} [2].
The splitting $\Delta m^2_{21}$, which drives the behavior of solar neutrinos, is roughly 30 times smaller than $\Delta m^2_{32} \cong \Delta m^2_{31}$, which drives the behavior of atmospheric neutrinos. (It is not known whether the closely-spaced pair $\nu_1$-$\nu_2$ is at the bottom or the top of the spectrum.) If an experiment is performed with $L/E$ such that $\Delta m^2_{32}\,L/E = {\cal O}(1)$, then  $\Delta m^2_{21}\,L/E \ll 1$, and in first approximation, this experiment cannot ``see'' the small splitting $\Delta m^2_{21}$. Neglecting this small splitting in \Eq{2.13}, this equation and the unitarity of $U$ imply that, for $\beta \neq \alpha$, 
\beeq
\mathrm{P}(\optbar{\na}\ra\optbar{\nb}) \cong 4|U_{\alpha 3}U_{\beta 3}|^2   \sin^2(\Delta m^2_{32} L/4E) ~~ .
\label{eq2.26}
\eeeq
Similarly, they imply that, for $\beta = \alpha$, 
\beeq
\mathrm{P}(\optbar{\na}\ra\optbar{\na}) \cong 1 - 4|U_{\alpha 3}|^2 (1-|U_{\alpha 3}|^2)  \sin^2(\Delta m^2_{32} L/4E) ~~ .
\label{eq2.27}
\eeeq
We see that, by measuring these simple oscillation probabilities, experiments with $\Delta m^2_{32}\, L/4E \\ = {\cal O}(1)$ can determine the flavor content of the isolated member of the spectrum, $\nu_3$.

\subsection{Neutrino oscillation in matter}{\label{s2.2}

Inside matter, the coherent forward scattering of neutrinos from the electrons, protons, and neutrons that make up the matter leads to neutrino effective masses and mixing angles that differ from their vacuum counterparts. As a result, inside matter, the probabilities for neutrino oscillations differ from their vacuum counterparts.

The Standard-Model interactions between neutrinos and other particles do not change flavor. Thus, barring hypothetical non-Standard-Model flavor-changing interactions, the observation of neutrino flavor change implies neutrino mass and leptonic mixing, even if the observation involves neutrinos passing through matter.

Neutrino propagation in matter may be conveniently treated via the laboratory-frame Schr\"{o}dinger time-evolution equation
\beeq
i \frac{\partial}{\partial t} \Psi(t) = {\cal H} \Psi(t) ~~ .
\label{eq2.28}
\eeeq
Here, $t$ is the time, and $\Psi(t)$ is a multi-component neutrino wave function. Its $\alpha$ component, $\Psi_\alpha (t)$, is the amplitude for the neutrino to have flavor $\alpha$ at time $t$. If there are $N$ flavors, the Hamiltonian ${\cal H}$ is an $N \times N$ matrix in flavor space. In matter, this matrix includes interaction energies arising from neutrino-matter interactions mediated by $W$ or $Z$ exchange. According to the Standard Model, the $Z$-mediated interactions neither change neutrino flavor nor depend on the flavor. Thus, they add to ${\cal H}$ a term proportional to the identity matrix. Such a term shifts all the eigenvalues of ${\cal H}$ by a common amount, leaving the splittings between the eigenvalues unchanged. Now, as we have seen when discussing neutrino flavor oscillation in vacuum, the amplitude for oscillation depends only on the {\em relative} phases of the different neutrino eigenstates. This means that it depends only on the splittings between the eigenvalues, and will not be affected by an interaction that merely shifts all the eigenvalues by the same amount. Thus, if our purpose is to treat neutrino flavor oscillation, we may omit the $Z$-exchange contribution to ${\cal H}$.

The $W$-exchange contribution is another matter. From the Standard Model, it follows that coherent forward $\nu_e$-electron scattering via the $W$-exchange diagram of Figure~\ref{f3} adds to the $\nu_e$-$\nu_e$ element of ${\cal H}$, ${\cal H}_{\nu_e \nu_e}$, an interaction energy
\beeq
V = \sqrt{2} G_F N_e ~~ .
\label{eq2.29}
\eeeq
\begin{figure}[!htbp]
\centering
\includegraphics[scale=1.00]{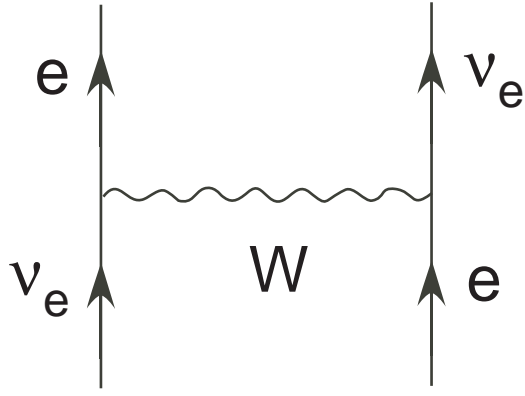}        
\caption{The $W$-exchange interaction that modifies neutrino flavor oscillation in matter.} 
\label{f3}
\end{figure}
Here, $G_F$ is the Fermi coupling constant, and $N_e$ is the number of electrons per unit volume in the matter through which the neutrinos are passing. The Fermi constant appears in $V$ because it is a measure of the amplitude for the diagram in Figure~\ref{f3}, and the density $N_e$ appears because the coherent scattering amplitude will obviously depend on how many electrons are present to contribute. The Standard Model tells us that for antineutrinos in matter, $V$ is replaced by $-V$.

Since $\nu_e$ is the only neutrino flavor that couples to an electron and a $W$, $W$-mediated $\nu - e$ scattering affects {\em only} the $\nu_e$-$\nu_e$ element of  ${\cal H}$. Thus, its contribution to ${\cal H}$ is not proportional to the identity matrix, and it does affect neutrino flavor oscillation.

Neutrino flavor change in matter is illustrated by the case where there are only two significant flavors, say $\nu_e$ and $\nu_\mu$, and, correspondingly, two significant mass eigenstates. The Hamiltonian ${\cal H}$ in the Schr\"{o}dinger equation, \Eq{2.28}, is then a $2\times 2$ matrix in $\nu_e$-$\nu_\mu$ space. Taking into account $\nu_e$-$e$ scattering via $W$ exchange, but omitting some irrelevant contributions that are proportional to the identity matrix, we readily find  [1] that
\beeq
{\cal H} = \frac{\Delta m^2_M}{4E}  \left[ \begin{array}{cc}
			 -\cos 2\theta_M  &  \sin 2\theta_M	\\
	\phantom{-}\sin 2\theta_M  &  \cos 2\theta_M
\end{array} \right] 	~~ .
\label{eq2.30}
\eeeq
Here, $E$ is the energy of the neutrinos, and $\Delta m^2_M$ and $\theta_M$ are, respectively, the effective (mass)$^2$ splitting and the effective mixing angle in matter. These effective quantities are related to their vacuum counterparts, $\Delta m^2$ and $\theta$, by
\beeq
\Delta m^2_M = \Delta m^2 \sqrt{\sin^2 2\theta + (\cos 2\theta - x_\nu)^2}
\label{eq2.31}
\eeeq
and
\beeq
\sin^2 2\theta_M = \frac{\sin^2 2\theta}{\sin^2 2\theta + (\cos 2\theta - x_\nu)^2} ~~ .
\label{eq2.32}
\eeeq
In these expressions,
\beeq
x_\nu \equiv \frac{2\sqrt{2} G_F N_e E}{\Delta m^2}
\label{eq2.33}
\eeeq
is a measure of the importance of the effects of matter. In vacuum, $N_e$ and consequently $x_\nu$ vanishes, and, as confirmed by Eqs.\ (\ref{eq2.31}) and (\ref{eq2.32}), $\Delta m^2_M$ and $\theta_M$ revert to the vacuum values, $\Delta m^2$ and $\theta$, respectively. 

Imagine an accelerator-generated neutrino beam that travels a distance $L \sim $1000\,km through the earth to a detector. The electron density $N_e$ encountered by this beam will be that of the earth's mantle, and approximately constant. Then $x_\nu,\; \Delta m^2_M ,\; \theta_M,\; $ and ${\cal H}$ will all be $\sim$\,position-independent. From Eqs.\ (\ref{eq2.30}) and (\ref{eq2.28}) and straightforward quantum mechanics, it follows that 
\beeq
\mathrm{P}(\nu_e \ra \nu_\mu)	 = \mathrm{P}(\nu_\mu \ra \nu_e) =   \sin^2 2\theta_M \sin^2(\Delta m^2_M L/4E) ~~ .
\label{eqI.4.1}
\eeeq
This is the usual two-neutrino oscillation result, \Eq{2.24}, except that the vacuum parameters $\theta$ and $\Delta m^2$ are replaced by their counterparts in matter, $\theta_M$ and $\Delta m^2_M$. If $N_e \ra 0$, so that $x_\nu \ra 0$, the oscillation probabilities in matter of \Eq{I.4.1} become the vacuum probabilities of \Eq{2.24}, as they must.

The size of the effect of matter may be judged by the size of $x_\nu$. For the illustrative beam that we are considering, the actual three-neutrino vacuum (mass)$^2$ splitting $\Delta m^2$ that will most strongly influence flavor oscillation is probably the large one, $\Delta m^2_{31} \cong \Delta m^2_{32}$. Experimentally, $\Delta m^2_{31} \simeq 2.4 \times 10^{-3}$\,eV$^2$ [8]. For this $\Delta m^2$, we find from \Eq{2.33} that 
\beeq
|x_\nu| \simeq E / 12\,\mathrm{GeV} ~~.
\label{eqI.4.2}
\eeeq
Thus, for $E = 0.5$\,GeV, the matter effect is quite small, for $E = 2$\,GeV it is modest, and for $E = 20$\,GeV it is large.

As already mentioned, when antineutrinos, rather than neutrinos, propagate through matter, the interaction energy $V$ is replaced by $-V$. It follows readily that, as a result, $x_\nu$, \Eq{2.33}, is replaced in Eqs.~(\ref{eq2.31})-(\ref{eq2.32}) by
\beeq
x_{\bar{\nu}}  \equiv -x_\nu ~~ .
\label{eq2.34}
\eeeq
We see that this change has the consequence that, within matter, the effective (mass)$^2$ splitting and the effective mixing angle for antineutrinos are different than they are for neutrinos. As a result, within matter the flavor oscillation of an antineutrino beam will differ from that of a neutrino beam. The two-flavor Hamiltonian ${\cal H}$ is still given by \Eq{2.30}, and the two-flavor oscillation probability in matter of constant density is still given by \Eq{I.4.1}, but the quantities $\Delta m^2_M$ and $\theta_M$ have different values than they did in the neutrino case.

Earlier, we raised the possibility that neutrinos are their own antiparticles. That is, we imagined that, {\em for a given momentum $\vec{p}$ and helicity $h$,} perhaps each neutrino mass eigenstate $\nu_i$ is identical to its antiparticle: $\nu_i(\vec{p},h) = \overline{\nu_i}(\vec{p},h)$. Suppose that this is indeed the case. Will the interaction with matter still cause the flavor oscillation of an ``antineutrino'' beam to differ from that of a ``neutrino'' beam within matter? In practical terms the answer is ``yes''. The reason is that, in practice, the ``neutrino'' and ``antineutrino'' beams that we study are never of the same helicity. A ``neutrino'' is the particle ``$\nu$'' produced, for example, in the decay $W^+ \ra e^+ + \nu$. As already noted, owing to the chirally left-handed structure of the weak interaction, this ``$\nu$'' will be of left-handed helicity: $h = -1/2$. In contrast, an ``antineutrino'' is the particle ``$\bar{\nu}$'' produced in $W^- \ra e^- + \bar{\nu}$. As already noted, owing to the same structure of the weak interaction, this ``$\bar{\nu}$'' will be of right-handed helicity: $h = +1/2$. Since the weak interaction is not invariant under parity, the interaction in matter of the left-handed ``$\nu$'' and the right-handed ``$\bar{\nu}$'' will be quite different, even if helicity is the only difference between the ``$\nu$''  and the ``$\bar{\nu}$''. Only the first term on the right-hand side of \Eq{2.1} can couple an incoming left-handed beam particle to an electron, while only the second term can couple an incoming right-handed beam particle. These two terms lead to different scattering amplitudes. These amplitudes do not depend on whether the ``$\nu$''  and ``$\bar{\nu}$'' beams differ only in helicity, or in some other way as well.

Future accelerator neutrino experiments hope to study $\nu_\mu \ra \nu_e$ and $\overline{\nu_\mu } \rightarrow \overline{\nu_e}$ in matter under conditions where all three of the known neutrino mass eigenstates $\nu_{1, 2, 3}$, or equivalently both of the known splittings $\Delta m^2_{31}$ and $\Delta m^2_{21}$, play significant roles. The oscillation probabilities are then more complicated than the expression of \Eq{I.4.1}. However, since $\alpha \equiv \Delta m^2_{21} / \Delta m^2_{31} \sim 1/30$ [2] and $\sin^2 2\theta_{13} < 0.2$ [9], the probability for $\nu_\mu \ra \nu_e$ in matter is well approximated by [10]
\beeq
P(\nu_\mu  \rightarrow \nu_e) \cong \sin^2 2\theta_{13}\,T_1 - \alpha \sin 2\theta_{13}\,T_2 +\alpha \sin 2\theta_{13}\,T_3 +  \alpha^2 T_4 ~~,
\label{eq2.40}
\eeeq
where
\beeq
T_1 = \sin^2 \theta_{23} \frac{\sin^2 [(1-x_\nu)\Delta]}{(1-x_\nu)^2} ~~ ,
\label{eq2.41}
\eeeq
\beeq
T_2 = \sin\delta \sin 2\theta_{12} \sin 2\theta_{23} \sin\Delta  \frac{\sin(x_\nu\Delta )}{x_\nu} \frac{\sin [(1-x_\nu)\Delta ]}{(1-x_\nu)} ~~ ,
\label{eq2.42}
\eeeq 
\beeq
T_3 = \cos\delta \sin 2\theta_{12} \sin 2\theta_{23} \cos\Delta \frac{\sin(x_\nu\Delta)}{x_\nu} \frac{\sin [(1-x_\nu)\Delta ]}{(1-x_\nu)} ~~ ,
\label{eq2.43}
\eeeq
and
\beeq
T_4 = \cos^2 \theta_{23}\sin^2 2\theta_{12} \frac{\sin^2 (x_\nu\Delta)}{x_\nu^2} ~~ .
\label{eq2.44}
\eeeq
In these expressions, $\Delta \equiv \Delta m^2_{31} L/4E$ is the kinematical phase of the oscillation. and $x_\nu$ is the matter-effect quantity defined by \Eq{2.33}, with $\Delta m^2$ now taken to be $\Delta m^2_{31}$.  In the appearance probability P($\nu_\mu  \rightarrow \nu_e$), the $T_1$ term represents the oscillation due to the splitting $\Delta m^2_{31}$, the $T_4$ term represents the oscillation due to the splitting $\Delta m^2_{21}$, and the $T_2$ and $T_3$ terms are the CP-violating and CP-conserving interference terms, respectively.
 
The probability for the corresponding antineutrino oscillation, P($\overline{\nu_\mu } \rightarrow \overline{\nu_e}$), is the same as the probability P($\nu_\mu  \rightarrow \nu_e$) given by 
Eqs.~(\ref{eq2.40})-(\ref{eq2.44}), but with $x_\nu$ replaced by $x_{\bar{\nu}} = -x_\nu$ and $\sin \delta$ by $-\sin \delta$: both the matter effect and CP violation lead to a difference 
between the $\nu_\mu  \rightarrow \nu_e$ and  $\overline{\nu_\mu } \rightarrow \overline{\nu_e}$ oscillation probabilities. In view of the dependence of $x_\nu$ on $\Delta m^2_{31}$, and in particular on the sign of $\Delta m^2_{31}$, the matter effect can reveal whether the neutrino mass spectrum has the closely-spaced $\nu_1$-$\nu_2$ pair at the bottom or the top (see Figure~\ref{f2}). 
However, to determine the nature of the spectrum, and to establish the presence of CP violation, it obviously will be necessary to disentangle the matter effect from CP violation in the neutrino-antineutrino oscillation probability difference that is actually observed. To this end, complementary measurements will be extremely important. These can take 
advantage of the differing dependences on the matter effect and on CP violation in 
P($\nu_\mu  \rightarrow \nu_e$) and P($\overline{\nu_\mu } \rightarrow \overline{\nu_e}$).

\section*{Acknowledgments}
It is a pleasure to thank H. Lipkin, S. Parke, and L. Stodolsky for useful conversations relevant to the physics of these lectures. I am grateful to Susan Kayser for her crucial role in the preparation of the manuscript.

\section*{References}
\frenchspacing
\begin{small}

\reference{1.} Kayser, B., ``Neutrino Physics'', in the{\it Proceedings of the SLAC Summer Institute of 2004}, eConf {\bf C040802}, L004 (2004): hep-ph/0506165.
\reference{2.} Kayser, B., ``Neutrino Mass, Mixing, and Flavor Change'', to appear in the 2008 edition of the {\it Review of Particle Physics}, by The Particle Data Group. This reference includes the phenomenology of neutrino oscillation and a summary of what we have learned about the neutrinos so far from experiment.
\reference{3.} This matrix is sometimes referred to as the Maki-Nakagawa-Sakata matrix, or as the Pontecorvo-Maki-Nakagawa-Sakata matrix, in recognition of the pioneering contributions of these scientists to the physics of mixing and oscillation. \\
See Maki, Z., Nakagawa, M., and  Sakata, S., {\it Prog.\ Theor.\ Phys.\ } {\bf 28}, 870 (1962); \\
Pontecorvo, B., {\it Zh.\ Eksp.\ Teor.\ Fiz.\ }{\bf 53}, 1717 (1967) [{\it Sov.\ Phys.\ JETP\ }{\bf 26}, 984 (1968)].
\reference{4.} For a discussion of the possibility of a nonunitary leptonic mixing matrix, 
see Antusch, S. {\it et al.}, {\it JHEP\ } {\bf 0610}, 084, (2006).
\reference{5.} Kayser, B., ``CP Effects When Neutrinos Are Their Own Antiparticles'', in {\it CP Violation}, ed. C. Jarlskog (World Scientific, Singapore, 1989) p. 334.
\reference{6.} Lipkin, H., \plb{642}{06}{366}.
\reference{7.} We thank S. Petcov for a long-ago conversation on how to obtain this result in a simple way.
\reference{8.} The MINOS Collaboration (Michael, D. {\it et al.}), \prl {97}{06}{191801}, and talks by MINOS collaboration members updating their results.
\reference{9.} The CHOOZ Collaboration (Apollonio, M. {\it et al.}),{\it  Eur.\ Phys.\ J.\ }{\bf C27}, 331 (2003); \\
Fogli, G. {\it et al.}, {\it Prog.\ Part.\ Nucl.\ Phys.\  }{\bf 57}, 742 (2006).
\reference{10.} Cervera, A. {\it et al.}, {\it Nucl.\ Phys.\ } {\bf B579}, 17 (2000); \\
Freund, M., \prd{64}{01}{053003}.

\end{small}
\end{document}